\documentclass[preprint]{JHEP} 
\usepackage{epsfig}
\usepackage{amsmath}
\usepackage{amssymb,amsfonts}

\newcommand{\bs}{\begin{subequations}}
\newcommand{\es}{\end{subequations}}
\newcommand{\nn}{\nonumber}

\newcommand{\ie}{i.e.\ }

\newcommand{\Zint}{\mathbb{Z}}
\newcommand{\Real}{\mathbb{R}}

\newenvironment{scalepic}[3]
  {\begin{center} 
   \begin{picture}(350,50)(-40,-25)%
   \put(-10,10){\makebox(0,0)[rb]{#1}}
   \put(-15,-10){\makebox(0,0)[rt]{#2}}
   \put(0,0){\vector(1,0){300}}
  }
  { \end{picture} \end{center} }
\newcommand{\scaleitem}[3]
 { \put(#1,0){ \put(0,-5){\line(0,1){10}}
               \put(0,10){\makebox(0,0)[b]{#2}}
               \put(0,-10){\makebox(0,0)[t]{#3}} }}

\title{On the short-distance structure of\\
irrational non-commutative gauge theories}

\author{Shmuel Elitzur\\
Racah Institute of Physics,
The Hebrew University, \\
Jerusalem 91904, Israel\\
E-mail: \email{elitzur@vms.huji.ac.il}}
\author{Boris Pioline\footnote{On
leave of absence from LPTHE, Universit{\'e} Pierre et Marie Curie,
PARIS VI and Universit{\'e} Denis Diderot, PARIS VII, Bo\^{\i}te
126, Tour 16, 1$^{\it er}$ {\'e}tage, 4 place Jussieu, F-75252
Paris CEDEX 05, FRANCE}\\
Jefferson Physical Laboratory, Harvard University\\
Cambridge, MA 02138, USA\\
E-mail: \email{pioline@physics.harvard.edu}}
\author{Eliezer Rabinovici\\
Racah Institute of Physics,
The Hebrew University, \\
Jerusalem 91904, Israel\\
E-mail: \email{eliezer@vms.huji.ac.il}}

\preprint{\hepth{0009009}\\
HUTP-00/A034\\RI-08-04}  

\abstract{As shown by Hashimoto and Itzhaki in {\tt hep-th/9911057}, 
the perturbative degrees of freedom of a non-commutative Yang-Mills theory 
(NCYM) on a torus are quasi-local only in a finite energy range.
Outside that range one may resort to a Morita equivalent 
(or T-dual) description appropriate for that energy. 
In this note, we study NCYM on a non-commutative torus
with an irrational deformation parameter $\theta$. 
In that case, an infinite tower of dual descriptions is generically needed 
in order to describe the UV regime. 
We construct a hierarchy of dual descriptions in terms of the
continued fraction approximations of $\theta$. We encounter
different descriptions depending on the level of the irrationality of
$\theta$ and the amount of non-locality tolerated.
The behavior turns out to be
isomorphic to that found for the phase structure of the four-dimensional 
Villain $\Zint_N$ lattice gauge theories, which we revisit as a warm-up.
At large 't Hooft coupling, using the AdS/CFT correspondance,
we find that there are domains of the radial coordinate $U$ where no T-dual
description makes the derivative expansion converge. The radial direction
obtains multifractal characteristics near the boundary of AdS.}


\begin{document}

\section{Introduction}
The non-commutativity introduced by the Moyal-Weyl star product
appears to give a consistent deformation of supersymmetric 
gauge theories \cite{ncdef,Connes:1998cr,nccons}, and to capture
part of the non-locality of open strings at least when they
propagate in non-trivial gauge 
backgrounds \cite{Connes:1998cr,Douglas:1998fm,Seiberg:1999vs,Schomerus}.
Whether it has any connection with the non-locality of quantum gravity
is unclear at this stage.
It nevertheless offers an interesting 
opportunity to study issues of non-locality in field theories.

A simple non-commutative manifold is the fuzzy torus, 
with a uniform non-commutativity scale $\Delta$.
Particles then behave as wave packets of Compton wavelength $1/E$ along their
direction of motion, but $E \Delta^2$ in the orthogonal direction. 
They can be interpreted as electric dipoles transverse to their 
motion \cite{dipole}.
For a rational deformation parameter $\theta=\Delta^2/\Sigma^2$,
where $\Sigma$ is the size of the box, 
non-commutativity does not imply non-locality at short distances however: 
it can be eliminated at the expense of 
enlarging the gauge group and introducing a magnetic flux.
This classical equivalence, known as Morita's \cite{Schwarz:1998qj,
tdual,morita}, is assumed
to hold true quantum mechanically for maximally 
supersymmetric gauge theories,
and descends directly from the T-duality of open strings \cite{tdual}.
This is not to say that the Moyal deformed gauge theory description
is altogether redundant:
instead, as explained in Ref. \cite{Hashimoto:1999yj},
it may serve in giving an appropriate quasilocal description 
of the degrees of freedom in a
finite energy range $1/\Sigma<E<1/(\theta\Sigma)$, in
which the Compton wave length and its dual are smaller than 
the size of the box. Beyond that range, one may find 
a Morita equivalent description, where the elementary 
particles are again localized on the scale of the box.
For $\theta$ rational, the dual 
3+1 dimensional commutative Yang-Mills theory then controls 
the ultraviolet behavior, while in the infrared the 
gauge theory undergoes a dimensional reduction to
1+1 dimensional commutative Yang-Mills. The intermediate 
energy range can be covered by a finite number of Morita dual 
patches (a single one in the simplest rational case of 
$\theta=1/s$) \cite{Hashimoto:1999yj}.

From the discussion above, the only situation where the non-locality
of the Moyal product may be effective is clearly when
the deformation parameter $\theta$ is irrational: the study of this case
is the primary goal of this note.
Before discussing the case of non-commutative Yang-Mills theories
with irrational $\theta$, a word is probably needed on the relevance
of such considerations for physics. Indeed, for all practical purposes,
any irrational number can always be approached by a rational number,
and phase transitions are not expected to occur in a dense
subspace of parameter space. Moreover, no way has yet been devised
to actually establish a measurable parameter as irrational.
Nevertheless, the difference between
rational and irrational can sometimes be hinted at by going to 
sufficiently high energy. The case of Villain $\Zint_N$ lattice gauge theories
\cite{lattice} in the presence of a $\theta$ angle (not to be confused with
the non-commutativity parameter) is particularly sharp in that respect: 
as the temperature increases, the system 
undergoes a sequence of phase transitions
where dyons of charge $(m,n)$ condense when $m/n$ approaches
$\theta$ \cite{Cardy}. When $\theta$ is irrational,
there is an infinite number of phase transitions as $T\to\infty$.
Of course, for all practical purposes, $T$ is bounded from above
and only a finite number of them are within reach. 
Other examples where the rationality of $\theta$ plays an important
role include the ``new'' decoupled six-dimensional gauge theories
\cite{6d},  and the non-commutative open string in 1+1 dimensions \cite{ncos}.
In these cases, the very definition of the model with irrational
$\theta$ angle or string coupling, respectively, is already problematic,
and it would be interesting to see if the methods used in this paper
are applicable.

In the case of Yang-Mills theories on a non-commutative torus,
the difference between rational and irrational
(and even between different kinds of irrational number) can 
be told by going to high energy. Indeed, for $\theta$ irrational,
it has been found that the cascade of Morita equivalent 
descriptions never terminates as the energy is 
increased \cite{Hashimoto:1999yj}. 
In this paper, we shall give an 
explicit construction of this cascade of pictures,
by relating it to the continued fraction approximation 
of $\theta$.\footnote{The relevance of continued fraction expansion
to non-commutative theories has been independently noticed 
in \cite{Landi:1999ey}, with different motivations.}
For $\theta$ rational, this will establish the existence
of a finite covering of the full energy range; for
$\theta$ irrational we will use it to study the
non-local UV behavior of the gauge theory. Depending on some properties of
irrational numbers and the level of confidence we require, we will see that 
one may or may not maintain a local description
at short distance. In particular, we will show that there exist values
of $\theta$ for which no quasi-local description exists beyond
a certain energy.
This is very similar to the case of $\Zint_N$ lattice gauge theories in
4 dimensions, where Coulomb phases occur when no dyon can condense,
and may prevail all the way to infinite temperature.

At strong 't Hooft coupling, we will find that a similar behavior holds:
for some irrational values of Neveu-Schwarz B-field, there exists
ranges of the radial coordinate for which no T-dual picture can give
a local (and convergent) supergravity description. In fact, we
shall argue that for $\theta$ irrational, the radial dimension 
becomes multifractal near the boundary of AdS. This is
an interesting infrared effect in gravity theories with irrational NS B-field,
which deserves further study.

The organization of this paper is as follow. In Section 2,
we shall take a look back at the problem of Villain $\Zint_N$ lattice gauge
theories, which will
allow us to introduce the relevant mathematics. The cascade
of Morita phases for NCSYM at weak coupling 
will then be constructed in Section 3. In Section 4, we shall
discuss the UV behavior in the large 't Hooft coupling regime,
from the point of view of the AdS/CFT correspondence,
and briefly discuss the situation at finite temperature.

\section{Phases of the Villain model}

As a warm up, we reconsider the phase structure of four dimensional
$\Zint_N$ lattice gauge theories \cite{lattice} in the presence of
a $\theta$ angle, following the discussion in \cite{Cardy}. 
These models are expected to capture universal features of
$U(N)$ Yang-Mills in the continuum. 
For simplicity, we consider a $\Zint_N$ lattice gauge
theory in the Villain approximation. The way to include a $\theta$
deformation reducing to the $\theta F\wedge F$ coupling in
the continuum limit was discussed in \cite{Cardy}. The result
is that the spectrum contains $(m,n)$ states with arbitrary magnetic
charge $m$ and electric charge $n$. 
Due to the $\theta$ angle deformation, the effective electric
charge $e=n-\theta m$ receives a fractional contribution from the magnetic
charge through the Witten effect \cite{Witten:1979ey}. 

The phase structure is controlled by the type
of states that condense in the vacuum: 
the Higgs phase corresponds
to a condensation of electric states with charge $(0,1)$ and hence
confinement of magnetic charges; the
ordinary confining phase corresponds to condensation of monopoles
with charge $(1,0)$. In addition, one may have 
oblique confinement whereby dyonic states
of charge $(m,n)$ condense \cite{tHooft:1981ht}. 
If no state condenses, the gauge fields
remain massless and we have a Coulomb phase. At weak coupling, or
small temperature, there is always a Higgs phase. The structure at
higher temperature depends however very sensitively on the value of $\theta$.
Indeed, a state with charge $(m,n)$ condenses if its free energy is less 
than that of the vacuum. The free energy was estimated in \cite{Cardy},
with the result that the condensation occurs if
\begin{equation}
\label{vil1}
\left( n-\theta m \right)^2 T + \frac{m^2}{T} < C/N\ , \quad T=N g^2/2\pi\ .
\end{equation}
Here $C$ is a numerical constant which can in principle be computed.
The l.h.s of this equation describes an ellipse
in the $(m,e=n-m\theta)$ 
plane whose aspect ratio depends on the temperature $T$.
The condensing states are the states of the lattice $\Zint\times \Zint$
contained in the ellipse and closest to the origin (see Figure 1a). 
For zero $\theta$
angle, the electric state $(1,0)$ clearly controls the low temperature 
behavior, while the magnetic state $(0,1)$ controls the high
temperature regime. The two phases may be connected 
for intermediate temperatures by a Coulomb phase
if $C/N$ is small enough so that the ellipse does not contain any
non-zero lattice point.

\EPSFIGURE{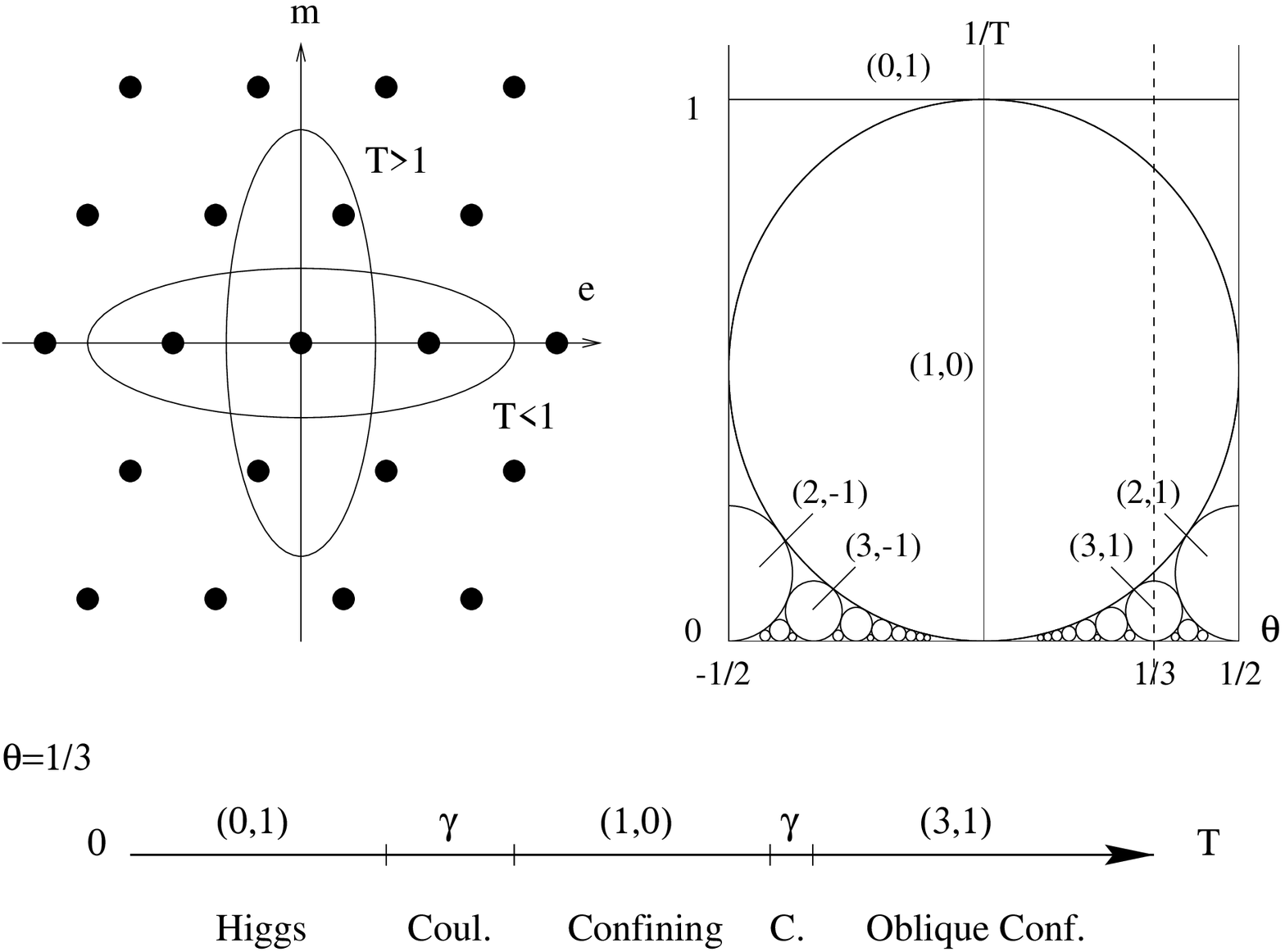,height=11cm}{Left(a): Condensation ellipse in the
$(m,e)$ plane. The state contained in the ellipse and closest to the origin
may condense at temperature $T$. Right(b): Condensation disks in the
$(\theta,1/T)$ plane for $M=2$. 
The disks denote the domain where a $(m,n)$ dyon 
may condense. Bottom(c): phase structure for $\theta=1/3$.}

The situation for non-zero but rational $\theta=p/q$ is 
already quite richer. At very high temperature, the ellipse 
will select the dyonic states with charge $(m,n)=(q,p)$,
which have zero effective electric charge $e=n-m\theta$:
the high temperature phase will therefore be an oblique
confining phase, in which the $(q,p)$ dyon condenses.
This will not occur however before $T>N q^2/C$.
Below this temperature, there will be other dyonic states
with small effective magnetic charge $e=n_{k}-m_{k}\theta$,
which will be able to condense at an intermediate temperature.
If $\theta$ is irrational, this will in fact be the case
at arbitrary high temperature: dyons with charge ration $n_k/m_k$
arbitrarily close to $\theta$ will condense successively,
and we will encounter an infinite number of phase transitions.

In more detail, there 
will exist a temperature $T$ at which the state $(m,n)$ will be able to
condense if the minimum value of the l.h.s in \eqref{vil1},
attained at $T=1/|\theta-n/m|$, is lower than the r.h.s, hence
\begin{equation}
\label{vil2}
| m ( n-\theta m)  |  < 1/M \ , \quad M=2N/C\ ,
\end{equation}
or equivalently
\begin{equation}
\label{vil3}
\left| \theta - \frac{n}{m} \right|  < \frac{1}{M m^2}\ .
\end{equation}
If this inequality holds, the condensation will occur in
a temperature interval
\begin{equation}
\label{vil4}
\frac{M m^2}{2 } < T < \frac{2}
{M (n-m\theta)^2 }
\end{equation}
where we assumed $|\theta-n/m| \ll 1$ for simplicity.

Evidently, the ability to approach $\theta$ by a rational number 
will determine the phase structure of the model.
Fortunately, the mathematical 
topic of Diophantine approximation
tells us a lot about the solutions of \eqref{vil3} (see
for instance \cite{cassels}). 
If $\theta=n/m$ is rational, it is known that there are only a finite number 
of fractions $n_k/m_k$ approching $\theta$ to order $1/(M m_k^2)$, irrespective
of the constant $M$ in \eqref{vil3}. Hence there will be a finite number
of phase transitions as $T$ increases from 0 to $\infty$.
In contrast, 
it is known that there is an infinite number of fractions $n/m$
approaching any irrational $\theta$ to order $1/(M m^2)$ with $M=1$ at least. 
The case $M=1$ is a simple application of Dirichlet's drawer principle,
which guarantees that for any integer $d$ there exists a fraction
$n/m$ with $m\leq d$ such that $|\theta-n/m|<1/(md)$.

There is in fact an algorithm allowing to obtain an infinite number
of solutions to \eqref{vil3}, namely the continued fraction decomposition. 
Recall that any irrational number can be obtained
as a limit of rational numbers $\theta_k$, constructed out a series
of positive integers $(c_k)$ through the continued 
fraction
\begin{equation}
\label{cfrac}
\theta_k=c_0 + \frac{1}{
\displaystyle c_1+ \frac{\strut 1}{
\displaystyle c_2+ \frac{\strut 1}{
\displaystyle\ddots{}~ c_{k-1}+ \frac{\strut 1}{c_k}}}} \ .
\end{equation}
The series $(c_k)$ specifies the real number $\theta$ uniquely. It truncates
when $\theta$ is rational, and becomes periodic if $\theta$ is a quadratic
irrational (\ie a root of a second order polynomial with integer 
coefficients). The series $\theta_k=n_k/m_k$ obtained
by truncating the fraction \eqref{cfrac} at $c_k$ converges
monotonically to $\theta$ from above for $k$ even, and from below for $k$ odd.
The integers $n_k$ and $m_k$ are strictly
increasing and can be computed using the recursion relations
\begin{eqnarray}
n_k&=&c_k n_{k-1}+n_{k-2}\ ,\quad n_0=c_0,\ n_1=c_0 c_1+1\ \nn\\
m_k&=&c_k m_{k-1}+m_{k-2}\ ,\quad m_0=1,\ m_1=c_1 \ \label{pqrec}
\end{eqnarray}
Note that the recursion can also be initiated by $(n_{-1},m_{-1})=(1,0)$.
These equations preserve the relation
\begin{equation}
\label{cop}
n_{k-1} m_{k}- n_{k} m_{k-1} = (-)^{k}
\end{equation}
so that the integers $(n_k,m_k)$ remain coprime.
The fractions $n_k/m_k$ are called 
the principal convergents of $\theta$, while the fractions
$n_{k}^{(p)}/m_k^{(p)}=(n_{k-2}+p n_{k-1})/(m_{k-2}+p m_{k-1})$
(with $p=1\dots c_k-1$) are called intermediate convergents 
and lie monotonically between $n_{k-2}/m_{k-2}$ and $n_k/m_k$.
The series $n_k/m_k$ can be shown to realize the condition \eqref{vil3} 
for $M=1$, \ie
\begin{equation}
\label{contappr}
\left|\theta- \frac{n_k}{m_k}\right| \leq \frac{1}{m_k^2}\ .
\end{equation}
In fact it can be shown that for any  pair of consecutive convergents
$n_k/m_k$ and $n_{k+1}/m_{k+1}$, at least one of the two satisfies
\eqref{vil3} for the stronger value $M=\sqrt{2}$; even better,
for any triplet  of consecutive convergents, at least one of the
three satisfies \eqref{vil3} for the stronger value $M=\sqrt{5}$. This value
cannot be improved, as one can show by taking $\theta$ to be
the Golden ratio $(1+\sqrt{5})/2$ (for which all $c_k=1$). If $\theta$ is
rational, the decomposition \eqref{cfrac} terminates, and the $n_k/m_k$
give the finite family of approximations to $\theta=n/m$ mentioned
in the previous paragraph. Using $1/(m_k\theta-n_k)^2\sim 
n_{k+1}^2/(m_k n_{k+1}-n_k m_{k+1})^2$ and \eqref{cop},
we see that the interval in \eqref{vil4} where the $(n_k,m_k)$
state condenses is immediately followed by the one where
$(n_{k+1},m_{k+1})$ condenses, so that the entire temperature range
is covered, by an infinite number of phases in the irrational
case.

Whereas we have obtained solutions to \eqref{vil3} with $M$ up to $\sqrt{5}$,
the value of the constant $C$ is not yet known, and may require 
finer rational approximations to $\theta$.
The ability to find solutions of \eqref{vil3} with  $M>\sqrt{5}$
depends crucially on the irrationality of $\theta$.
To be more precise, let $M(\theta)$ define the supremum
of all $M\in \Real^+$ such that there exist an infinity of
fractions $n/m$ approching $\theta$ closer than $1/(M m^2)$.
Note that $M(\theta)$ is invariant under fractional linear transformations
$\theta\to (a\theta+b)/(c\theta+d)$.
It can in fact be shown that $M(\theta)=\sqrt{5}$ if and only if
$\theta$ is equivalent to
the Golden Ratio $\theta_1=(1+\sqrt{5})/2$ up to a $Sl(2,\Zint)$
transformation. If not, then $M(\theta)
\geq 2\sqrt{2}$,
with equality iff $\theta$ is equivalent to $\theta_2=1+\sqrt{2}$.
More generally, there exists an infinite series $(\theta_k)$ with
increasing 
$M(\theta_n)\to 3$, such that any $\theta$ with $M(\theta)<3$ is equivalent
to one of the $\theta_n$'s. These numbers $\theta_n$ form the Markoff chain,
and are quadratic algebraic numbers, roots of second order polynomials
which can be obtained in a recursive way \cite{cassels}.
Finally, most irrational numbers satisfy $M(\theta)\geq 3$,
with infinitely many such that $M(\theta)=3$. 
This state of affairs is represented on Figure 2.
Coming back to the physical problem at hand, 
if $C/(2N) > 1/M(\theta)$, there will be an infinity
of states condensing as the temperature is increased from
0 to $\infty$. If however 
$C/(2N) < 1/M(\theta)$, as always occurs at large enough $N$, there will only
be a finite number of phase transitions. This
means that the system will reach a Coulomb phase at high temperature,
after passing a finite number of phase transitions.
This is a rather unusual case where, even at infinite temperature,
the symmetry is not restored. The system behaves at high temperature
as if the gauge group was non-compact, and thus did not possess topological
excitations. The reader should however keep in mind that this statement
holds within the Villain approximation.

\begin{figure}[t]
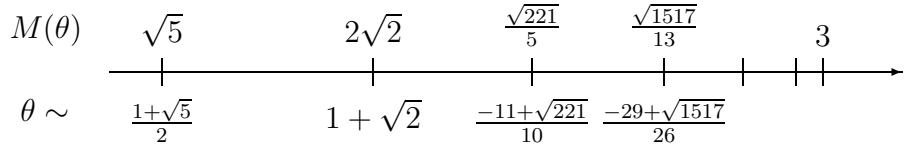

\begin{scalepic}{$M(\theta)$}{$\theta\sim$}{scal0}
\scaleitem{20}{$\sqrt{5}$}{$\frac{1+\sqrt{5}}{2}$}
\scaleitem{100}{$2\sqrt{2}$}{$1+\sqrt{2}$}
\scaleitem{160}{$\frac{\sqrt{221}}{5}$}{$\frac{-11+\sqrt{221}}{10}$}
\scaleitem{210}{$\frac{\sqrt{1517}}{13}$}{$\frac{-29+\sqrt{1517}}{26}$}
\scaleitem{240}{}{}
\scaleitem{260}{}{}
\scaleitem{270}{$3$}{}
\end{scalepic}
\caption{Maximum $M$ achievable in \eqref{vil3} while preserving
an infinity of solutions. $M(\theta)\geq 3$ except for a discrete
series of quadratic irrationals $\theta_n$.\label{mtheta}}
\end{figure}

Finally, we would like to mention an alternate way of understanding
the phase structure of the Abelian model. 
We first note following \cite{Cardy}
that the free-energy in \eqref{vil1} is invariant under electric-magnetic
duality, acting by $Sl(2,\Zint)$ fractional linear transformations on
the complexified coupling $\tau=\theta+ i/T$,
\begin{equation}
\label{sl2}
\tau\to\frac{c+d\tau}{a+b\tau}\ ,\quad
\left(\begin{array}{c} m \\ n \end{array}\right)
\to \left(\begin{array}{cc} a & b \\ c & d \end{array}\right)
\left( \begin{array}{c} m \\ n \end{array}\right)
\ ,\quad ad-bc=1\ .
\end{equation}
The temperature and $\theta$ angle can therefore always be mapped
back into the fundamental domain of the upper half plane,
$|\theta|<1/2, |\tau|>1$. For $\theta=0$, we need only two
patches $T>1$ and $T<1$ to cover the semi-infinite line $T>0$,
and the Higgs and confining phases mentioned above are simply
related to each other by duality $T\to 1/T$. For a
fixed non-zero rational $\theta$, it takes finitely many
patches to cover the line $\tau=\theta+i \Real^+$, and hence
we expect finitely many phase transitions. For an irrational
$\theta$, it takes an infinite number of patches, and hence we 
expect infinitely many phase transitions as $T\to\infty$.
Each $(m,n)$ dyon  defines a disk on the upper half place,
centered at $\tau=n/m+i/(M m^2)$ and radius $1/(M m^2)$
(hence tangent to the real axis) 
which delimits the region of temperature and $\theta$
angle where it might condense (see Fig. 1, right). The space
between the disks correspond to Coulomb phases.
For the critical value $M=2$, all disks are tangent to each
other. Following a line $\tau=\theta+i \Real^+$ from infinity,
one encounters an infinite number of phase transitions, except
when $\theta$ is rational. For $M>2$, the disks do not touch
each other, and one may find a line such which does not encounter
any disk as $T\to\infty$. For $M<2$, the disks start overlap,
in which case it is the state with the lowest free energy 
that condenses. For $M<\sqrt{3}$,
the disks cover all the upper-half-plane, and there are no Coulomb
phases any more.

Having used the Villain model as a setting to introduce the
mathematical properties we need, we now come to the
proper topic of this note, namely gauge theories
on a non-commutative torus.

\section{NCSYM on an irrational torus at weak coupling}
Following \cite{Hashimoto:1999yj}, let us consider the $N=4$ 
supersymmetric $U(N)$ gauge theory on $\Real^2\times T^2_{\theta}$
where $T^2_{\theta}$ is a non-commutative torus with non-locality
scale $\Delta$ and radius $\Sigma$. The dimensionless
deformation parameter is $\theta=\Delta^2/\Sigma^2$.
We also allow for a magnetic background $m$ on $T^2$.
As we recalled in the introduction, 
the non-commutative gauge theory gives a quasi-local description 
at energies 
\begin{equation} 
\frac{1}{\Sigma} < E < 
\frac{\Sigma}{\Delta^2}= \frac{1}{\theta\Sigma}
\label{erange}
\end{equation}
At either end of this interval, 
the wave function starts spreading
over the whole torus $T^2$, either in the direction of motion
(at the lower end) or in the direction transverse to it (at the upper
end). In the lower half $E<1/\Delta$ of the interval \eqref{erange}, 
the effect of the non-commutativity is not noticeable, but this
does not affect the statement of quasi-locality and we will not
elaborate on this point here.
These finite size effects may be avoided by looking for a Morita dual
picture which brings the energy $E$ within the new quasi-local
window $1/\tilde \Sigma<E<1/\tilde\theta\tilde\Sigma$.
An $Sl(2,\Zint)$ Morita transformation acts as
\begin{eqnarray}\label{ol}
&&\tilde \theta = \frac{c + d \theta}{a + b \theta}, \qquad 
\tilde \Phi = (a + b \theta)^2 \Phi - b ( a + b \theta), \qquad 
\tilde{\Sigma} = (a + b \theta) \Sigma, \nonumber \\
&&\tilde{g}_{YM}^2 = (a + b \theta) g_{YM}^2, \qquad 
\left(\begin{array}{c} \tilde m \\ \tilde N \end{array}\right)
= \left(\begin{array}{cc} a & b \\ c & d \end{array}\right)
\left( \begin{array}{c} m \\ N \end{array}\right),
\end{eqnarray}
where $\Phi$ is the standard magnetic background \cite{Schwarz:1998qj,morita}. 
We thus have to look for integers $a,b,c,d$ such $ad-bc=1$ and
\begin{equation} 
\label{erange2}
\frac{1}{|a+b\theta| \Sigma} < E < \frac{1}{|c+d\theta|\Sigma}
\end{equation}
If $\theta$ is rational, we may choose $\theta=-c/d$ such that
the upper bound is infinite. The quasi-local description 
in the UV is therefore
an ordinary commutative gauge theory, albeit with a magnetic
flux. If $\theta$ is irrational,
the upper bound can never be made infinite, and we
have to keep finding dual descriptions as we increase the energy.

Given a fixed energy $E$, we may ask what is the optimum duality
picture which allows to have a quasi-local description in the
largest interval possible around $E$. In order to optimize the upper bound,
we choose $|\theta+c/d|\ll 1$. Using $ad-bc=1$, we can rewrite the inequality
\eqref{erange2} as
\begin{equation}
\label{erange3} 
d < E \Sigma < \frac{1}{|c+d\theta|}
\end{equation}
This is equivalent to
\begin{equation} 
\label{minmax}
|\theta+c/d| < \frac{1}{d~E\Sigma}\ ,\quad d \leq E\Sigma
\end{equation}
By the same Dirichlet's drawer principle mentioned in Section 1, 
it is therefore clear that for any $E$ there
exists a fraction $c/d$ such that the above condition  is verified.
Note that the two conditions in \eqref{minmax} imply 
\begin{equation} 
|\theta+c/d| < \frac{1}{d^2}
\end{equation}
which does not make any reference to the energy any more.
In particular, the continued fraction convergents of $\theta$ fulfill
this condition but it is easy to see that they are not the 
only ones. The principal and intermediate convergents of $\theta$
however have the property of being
{\it best approximations} to a given rational or irrational $\theta$,
in the sense that any closer fraction has to have a bigger denominator.
By choosing $c/d$ as a convergent of $\theta$, we thus minimize
the lower bound of \eqref{minmax} while maximizing the higher bound.
Note that $a+b\theta$ tends to zero at the same time as $c+d\theta$ does,
so that the Yang-Mills coupling $\tilde g_{YM}$ keeps decreasing as we
go to the UV. The 't Hooft coupling $\lambda=g_{YM}^2 (N-\theta m)$
however remains unchanged under
Morita transformations.

Having justified the relevance of continued fraction approximations,
we now proceed to show that the series of convergents $n_k/m_k$
provides a complete covering of the full energy range.
From the relation \eqref{cop}, we see that the linear fractional transformation
\begin{equation}
\theta\to (-)^{n-1} \frac{m_k \theta- n_k}{m_{k-1} \theta - n_{k-1}}
\end{equation}
is a bona fide $Sl(2,\Zint)$ Morita transformation.
Using the formulae \eqref{ol}, we find that the new range of validity
of this dual picture
\begin{equation} 
\label{erange4}
\frac{1}{|n_{k-1}-m_{k-1}\theta| \Sigma} 
< E < \frac{1}{|n_{k}-m_{k}\theta|\Sigma}\ .
\end{equation}
Since $n_k-m_k\theta\to 0$ as $k$ goes to infinity, we see that we have
succeeded in covering the entire energy range with 
non-overlapping dual descriptions, for any irrational $\theta$.
In analogy with our discussion of the Villain model, one may want
to introduce a confidence level in the  criterion
\eqref{erange}, and ask for which irrational values of $\theta$
the cascade of quasi-local descriptions
can still be constructed. We will address this question in the
next section, after discussing the strong coupling version of the
criterion \eqref{erange}. The results will be the same, as far as the $\theta$ 
dependence is concerned, as those to be uncovered at strong coupling.

\section{Strongly coupled NCSYM and multifractal gravity dual}
We now turn to the  strong 't Hooft
coupling of NCSYM on a torus with irrational $\theta$, 
which we can study using the AdS/CFT correspondence.
We assume the 't Hooft coupling $\lambda$ and $GCD(N,m)$ to be large.
The metric, dilaton and B-field dual to $N=4$ SYM
on a non-commutative torus have been obtained in \cite{Hashimoto:1999ut,
Maldacena:1999mh}, and read
\begin{eqnarray}
\label{rt}
ds^2&=& l_s^2 \left\{
\frac{U^2}{\sqrt{\lambda}}(-dt^2+dx_1^2) +\frac{\sqrt{\lambda}
U^2}{\lambda + U^4 \Delta^4}(dx_2^2+dx_3^2)
+\frac{\sqrt{\lambda}}{U^2}dU^2 +\sqrt{\lambda}d \Omega_5^2 \right\},
\nonumber \\ 
e^{\phi}&=&\frac{\lambda}{4\pi N}
\sqrt{\frac{\lambda}{\lambda + \Delta^4 U^4}}, \qquad
B_{23} = -\frac{l_s^2 \Delta^2 U^4}{\lambda + \Delta^4 U^4}, \label{sugra}
\end{eqnarray} 
with periodicities $x_2 \sim x_2 +2 \pi \Sigma$ and $x_3 \sim x_3 +2\pi\Sigma$.
This gravity solution is appropriate for vanishing magnetic background
$m=0$, which we assume from now on. This is no loss of generality,
since one may always choose $m=0$ and $|\theta|<1/2$ through
a duality transformation \cite{Barbon}. 

The supergravity can be trusted when the volume of the torus is 
bigger than the string scale,  hence
\begin{equation}
\label{vu}
v=\frac{\sqrt{\lambda}U^2\Sigma^2}{\lambda+U^4\Delta^4} > \frac{M}{2}
\end{equation}
where the r.h.s. is the level of confidence we wish to impose on
the supergravity description. If this condition is not satisfied,
the Kaluza-Klein modes of the supergravity fields 
mix with the string winding states, and a supergravity description is
no more sufficient. One may be tempted to identify $M$ with the radius
of convergence of the $\alpha'$ expansion, but the derivative expansion
is presumably only asymptotic, so that this identification cannot be
precise. We shall keep with the somewhat loose notion of ``confidence level''
in the following, although a better understanding of the constant $M$
would certainly be desirable.

In terms of the dimensionless variable $u=U\Sigma/\lambda^{1/4}$, the volume
of the torus in string units becomes
\begin{equation}
v=u^2/(1+\theta^2 u^4) 
\end{equation}
which attains its maximum value $1/(2|\theta|)$
at $u=1/\sqrt{|\theta|}$. The torus is therefore bigger than
the string length for $|\theta|<1/2$, in the range $u_-<u<u_+$
with $u_{\pm}^2=(1\pm \sqrt{1-M^2\theta^2})/(M \theta^2)$.
If $\theta\ll 1$, the supergravity solution can therefore be 
trusted in the range
\begin{equation}
\label{urange}
\frac{\lambda^{1/4}}{\Sigma}
\sqrt\frac{M}{2}< U < \frac{\lambda^{1/4} \Sigma}{\Delta^2}
\sqrt\frac{2}{M}
\end{equation}
with maximum reliability at $U=\lambda^{1/4}/\Delta$.
Using the AdS/CFT dictionary and choosing $M=2$, 
this translates into an energy range
\begin{equation}
\frac{1}{\lambda^{1/4}\Sigma}< E < \frac{\Sigma}{\lambda^{1/4}\Delta^2}
\end{equation}
which agrees with the field theory criterion \eqref{erange} up to a factor
of $\lambda^{1/4}$, as observed in \cite{Hashimoto:1999yj}.
The requirement of quasi-locality at strong 't Hooft coupling
(ie neglecting the non-local string winding states around the gravity
background) therefore matches the requirement of quasi-locality 
at weak coupling, up to the usual large-N renormalisation
by the factor of 't Hooft coupling. Note that the confidence level
$M$ could also have been introduced in the weak coupling criterion
\eqref{erange} by specifying the size of the energy window.

Outside this energy range, we may still find a reliable 
supergravity description
by going to a T-dual picture (see \cite{Giveon:1994fu} for a review). 
Applying an $Sl(2,\Zint)$ T-duality 
transformation, 
\begin{equation}
\label{tdual}
\rho\to\frac{a\rho+b}{c\rho+d}\ ,\quad \rho=\frac{\Sigma^2}{l_s^2}
B_{23}+ i v\ ,\quad v e^{-2\phi}={\rm const.}
\end{equation}
the volume of the dual torus in string units and the
dilaton become
\begin{equation}
\label{vucd}
v=  \frac{u^2}{d^2+ (d\theta-c)^2 u^4} \ ,
\quad
e^{\phi}=\frac{\lambda}{4\pi N}\frac{1}{\sqrt{d^2+(d\theta-c)^2 u^4}}\ .
\end{equation}
In particular, the string coupling remains smaller than
the one in the original picture, at least by a factor $1/d$.
Requiring again the volume to be bigger than a fixed constant $M/2$,
we obtain a condition
\begin{equation}
\frac{d^2}{u^2}+ (d\theta-c)^2 u^2 < \frac{2}{M}
\end{equation}
isomorphic to the condition \eqref{vil1} appearing in the Villain model,
under the identification
\begin{equation}
\tau=\theta+i/T \ \mbox{(Villain)} \quad\leftrightarrow\quad
\tau=\theta+ i/u^2 \ \mbox{(AdS)} 
\end{equation}
The constant-$\theta$ vertical line in the phase diagram of the
Villain model (Figure 1b) is therefore identified with
the radial direction of AdS for a fixed value of the NS B-field.
The disks now indicate what is the appropriate T-dual
description (if any) for any point along the $u$ axis.
From \eqref{rt}, the metric along the radial direction is $(du/u)^2$,
so that the phase structure is better represented in the
$\theta+i \ln(u)$ plane, which corresponds to proper distance in 
target space (Figure 3).
\EPSFIGURE{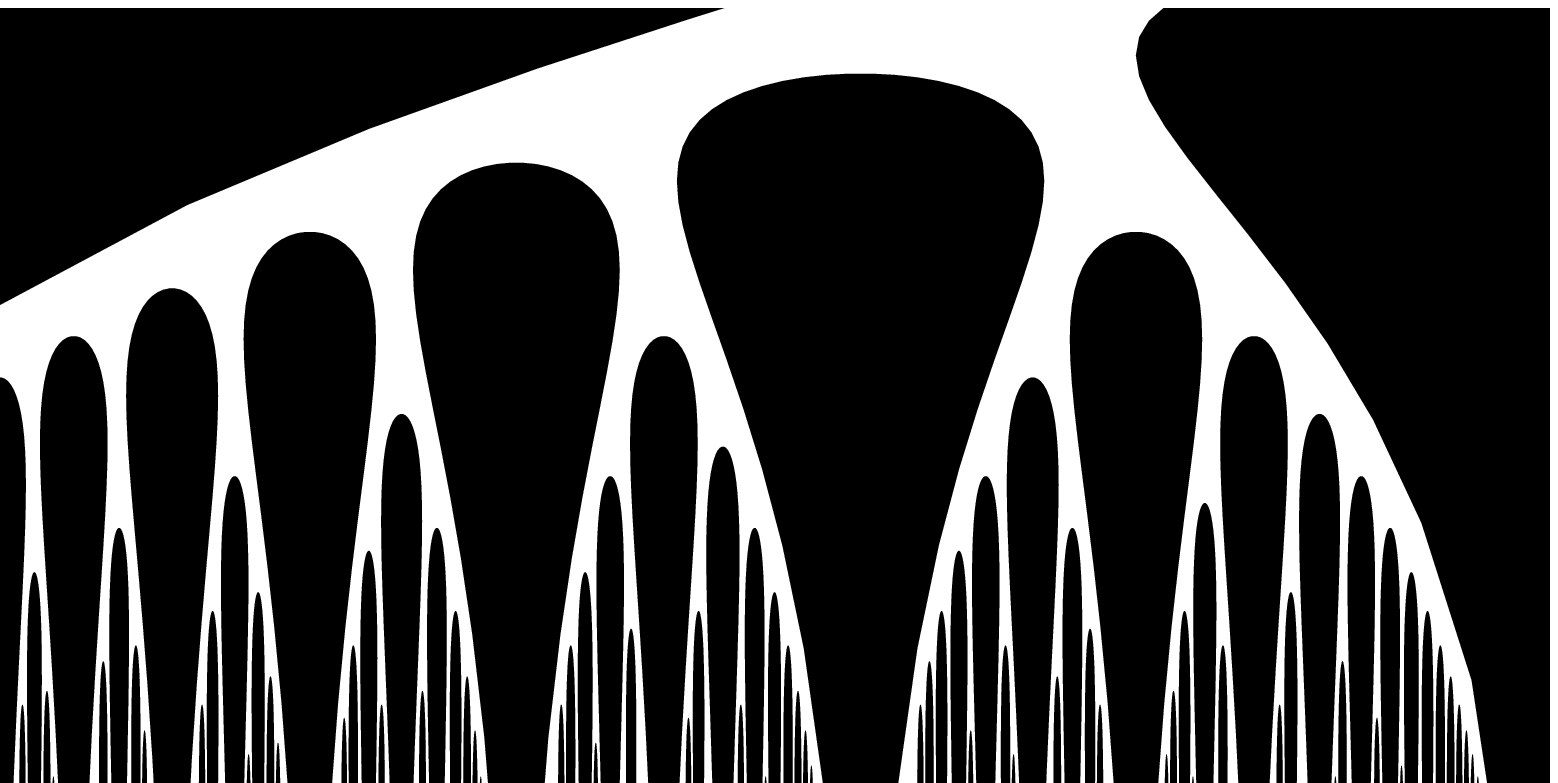,height=7cm}{Geometry of the radial dimension $U$ 
(measured in proper units) with respect to the NS B-field. The shaded area
is where a dual local supergravity description holds (we chose $M=2.5$ for 
clarity of the picture). The area displayed
is $1/8<\theta<1/2$ and $-7<\ln (1/u^2)<-2$.}

From this picture, we see that the interior of AdS ($u\to 0$, i.e.
the infrared regime of the gauge theory) is universally 
described by the $(c,d)=(1,0)$ picture, where the volume of 
the torus diverges at $u\to 0$. The system is then describable
by a purely D1-brane system (\ie $N=0$), and undergoes a Gregory-Laflamme
transition \cite{GL}: 
this is the standard phase transition from 3+1 NCSYM to 1+1 
SYM \cite{Kogan}.
In contrast, at the boundary of
AdS  ($u\to\infty$, \ie to the UV of the field theory), the behavior
depends crucially on the irrationality of $\theta$.
For any rational value of $\theta$, 
the radial direction ends up into one of the shaded areas (\ie one 
of the cusps at the bottom of Figure 3), and remains there
all the way towards the boundary of AdS. However, for $\theta$ irrational,
the radial direction keeps switching from one shaded area to
another. If $M>2$, the shaded areas are not connected, and
there exists values of $\theta$ for which there is no
reliable description in some radial range.
If $M<\sqrt{3}$, the complete $\tau$ plane is covered, and there
is always a trustable T-dual description.
Depending on the value of $M(\theta)$, the domain of radial positions for which
a reliable description exists may be finite ($M(\theta)<M$) or infinite
($M(\theta)>M$). Note that the string coupling is not an issue,
since $e^{\phi}$ decreases to zero as $d$ increases to infinity.
As in \eqref{vil2} and \eqref{vil3}, the
dual supergravity description has a non-empty range of radial
positions in which it can be trusted
if the minimum of the volume $v$ is larger than $M/2$, \ie
\begin{equation}
\frac{1}{2|d(d\theta-c)|} > \frac{M}{2}\ .
\end{equation}
The range of validity is then 
\begin{equation}
\label{urange2}
|d|\sqrt\frac{M}{2}
< \lambda^{1/4} U\Sigma< \frac{1}{|d\theta-c|}\sqrt\frac{2}{M}
\end{equation}
The latter condition can be rewritten as $|\theta-c/d|<1/(M d^2)$,
which admits a finite ($M(\theta)<M$) or infinite
($M(\theta)>M$) number of solutions.
The condition \eqref{urange2} for $M=2$ agrees with the field theory condition
\eqref{erange} up to the coupling factor $\lambda^{1/4}$, upon noting
that $d\sim 1/(a+b\theta)$ if $|\theta-c/d|\ll 1$. 

Quite strikingly, the picture 
reveals the fractal nature of the radial dimension
for irrational $\theta$ at the boundary of AdS. 
For $\theta$ a quadratic irrational, the continued fraction 
decomposition is periodic, and so is the phase structure on
the radial direction (in logarithmic units): the radial direction has therefore
a definite fractal dimension, related to the periodicity
of the expansion. For irrational numbers that are not quadractic
however, the ratio of the energies at the branching 
points is quite erratic, and the radial dimension turns 
out to be multifractal (see \cite{Mandelbrot} for a review). 
These two cases are illustrated on Figure 4.
If $M>M(\theta)$, the radial direction 
is completely stringy towards the boundary of AdS, and the dimension of the
radial direction covered by duality pictures is therefore 0.
It would be very interesting to compute the Hausdorff dimension and
the spectrum of fractal exponents for general $\theta$
in terms of $M$ and $M(\theta)$. As a word of caution, we emphasize that
the fractality of the radial dimension only arises towards the
boundary of AdS, \ie as an infrared effect in the gravity theory.
Whether non-commutativity can also be relevant for the description
of the ultraviolet behavior of quantum gravity remains to be seen.
\FIGURE{
\epsfig{file=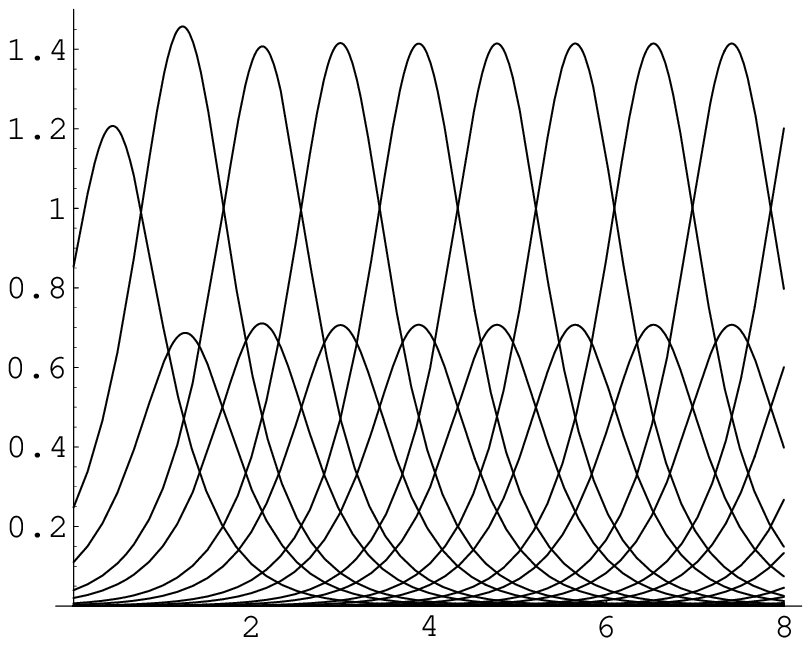,height=5cm}\epsfig{file=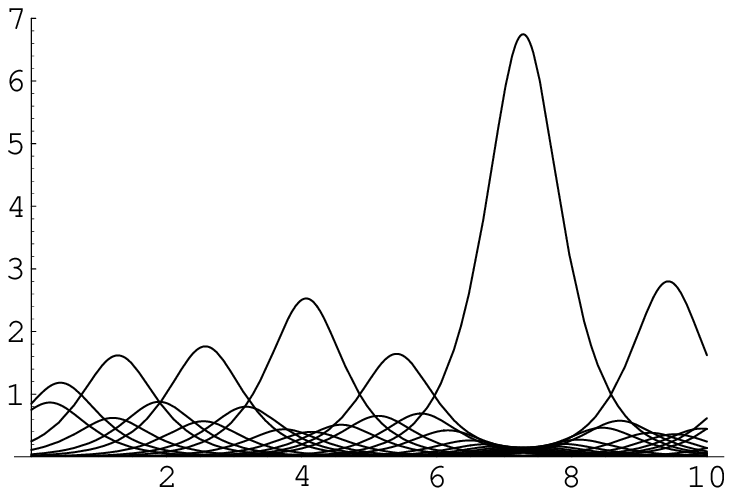,height=5cm}
\caption{Volume of $T^2$ in string units versus radial
distance in logarithmic units for dual pictures given by the 
convergents and intermediate convergents of $\theta=\sqrt{2}-1$ (left)
and $\theta=\gamma_E\sim .577216$ (right).}
}

We conclude with three further comments on the phase structure of NCSYM.
In the situation we have considered, there is no phase transition
between different patches, but only a cross-over in the high
energy behavior. One may ask whether this picture changes when
we put the gauge theory on a non-commutative torus at finite temperature.
The gravity
dual was obtained in \cite{Maldacena:1999mh}, and differs from \eqref{rt}
in the $g_{tt}$ and $g_{rr}$ components only, which
are affected by a factor $(1-U_0^4/U^4)$. The 
horizon at $U=U_0$ effectively cuts off the
higher part of the diagram in Figure 1b or Figure 3,
but leaves the ultraviolet behavior unchanged.
The thermodynamic quantities are invariant
under T-duality, so that we do not expect any phase
transitions.
Second, we may ask whether a Gregory-Laflamme transition can
occur in the interior of AdS. This would in principle be
achievable by tuning the ratio $N/m$ close to $\theta$.
However, one may show that such a patch with only D1-branes
($N=0$) always extends to $u\to 0$ \cite{Barbon}. 
The Gregory-Laflamme transition 
can therefore only occur in the infrared of the gauge theory.
Finally, we should emphasize that the issue of quasi-locality
concerns only the effective degrees of freedom at a given energy scale:
local operators have smooth correlation functions at all scales,
and may in fact become non-local in the dual pictures that we
introduced at higher energy.

\section*{Acknowledgments}
The authors are grateful to J.L.F. Barb\'on and N. Seiberg  for
valuable discussions and suggestions. 
B.P. is indebted to N. Itzhaki for several discussions in Nov. 1999
on the UV behaviour of non-commutative gauge theories.
E.R. is grateful to S. Shenker and S. Shenker for educating him
on number theory in 1981, and pointing out its relevance to the Villain 
model, respectively. 
Hospitality and support of the Racah Institute of
Physics in Jerusalem, the CERN Theory Division and
the Tokyo Institute of Technology, Meguro, Tokyo
152-8551, extended to B.P. and
E.R. respectively during part of this work
are gratefully acknowledged. 
This research is supported in part by the European
RTN network HPRN-CT-2000-00122,
the David and Lucile Packard foundation, the BSF-American-Israeli Bi-National
Science Foundation, the IRF Centers of Excellence Program
and the Japan Society for
the promotion of sciences.


\providecommand{\href}[2]{#2}
\begingroup\raggedright\endgroup


\begin{thebibliography}{10}

\bibitem{ncdef}
A.~Gonzalez-Arroyo and C.~P.~Korthals Altes,
``Reduced Model For Large N Continuum Field Theories,''
Phys.\ Lett.\  {\bf B131} (1983) 396.
A.~Connes and M.~Rieffel {\it Yang-Mills for 
noncommutative two-tori}, Operator algebras and Mathematical 
Physics (Iova City, Iova, 1985), pp. 237-266, Contemp. Math. Oper. Algebra. 
Math. Phys. 62, AMS 1987. 


\bibitem{Connes:1998cr}
A.~Connes, M.~R.~Douglas and A.~Schwarz,
``Noncommutative geometry and matrix theory: Compactification on tori,''
JHEP {\bf 9802}, 003 (1998)
[hep-th/9711162].

\bibitem{nccons}
T.~Filk,
``Divergencies in a field theory on quantum space,''
Phys.\ Lett.\  {\bf B376} (1996) 53.
J.~C.~Varilly and J.~M.~Gracia-Bondia,
``On the ultraviolet behaviour of quantum fields over noncommutative  manifolds,''
Int.\ J.\ Mod.\ Phys.\  {\bf A14}, 1305 (1999)
[hep-th/9804001].
M.~M.~Sheikh-Jabbari,
``Renormalizability of the supersymmetric Yang-Mills theories on the  noncommutative torus,''
JHEP {\bf 9906}, 015 (1999)
[hep-th/9903107].
C.~P.~Martin and D.~Sanchez-Ruiz,
``The one-loop UV divergent structure of U(1) Yang-Mills theory on  noncommutative $R^4$,''
Phys.\ Rev.\ Lett.\  {\bf 83}, 476 (1999)
[hep-th/9903077].
T.~Krajewski and R.~Wulkenhaar,
``Perturbative quantum gauge fields on the noncommutative torus,''
Int.\ J.\ Mod.\ Phys.\  {\bf A15}, 1011 (2000)
[hep-th/9903187].




\bibitem{Douglas:1998fm}
M.~R.~Douglas and C.~Hull,
``D-branes and the noncommutative torus,''
JHEP {\bf 9802}, 008 (1998)
[hep-th/9711165].



\bibitem{Seiberg:1999vs}
N.~Seiberg and E.~Witten,
``String theory and noncommutative geometry,''
JHEP {\bf 9909}, 032 (1999)
[hep-th/9908142].

\bibitem{Schomerus}
V.~Schomerus,
JHEP {\bf 9906}, 030 (1999)
[hep-th/9903205].

\bibitem{dipole}
M.~M.~Sheikh-Jabbari,
``Open strings in a B-field background as electric dipoles,''
Phys.\ Lett.\  {\bf B455}, 129 (1999)
[hep-th/9901080];
D.~Bigatti and L.~Susskind,
``Magnetic fields, branes and noncommutative geometry,''
hep-th/9908056.

\bibitem{Schwarz:1998qj}
A.~Schwarz,
``Morita equivalence and duality,''
Nucl.\ Phys.\  {\bf B534}, 720 (1998)
[hep-th/9805034].
M.~Rieffel and A.~Schwarz, {\it Morita equivalence of 
multidimensional noncommutative tori}, q-alg/9803057.


\bibitem{tdual}
P.-M. Ho, ``Twisted bundle on quantum torus and BPS states in matrix theory,''
  {\em Phys. Lett.} {\bf B434} (1998) 41--47,
  \href{http://xxx.lanl.gov/abs/hep-th/9803166}{{\tt hep-th/9803166}};
D.~Bigatti, ``Non commutative geometry and super Yang-Mills theory,'' {\em
  Phys. Lett.} {\bf B451} (1999) 324,
  \href{http://xxx.lanl.gov/abs/hep-th/9804120}{{\tt hep-th/9804120}};
G.~Landi, F.~Lizzi, and R.~J. Szabo, ``String geometry and the noncommutative
  torus,'' {\em Commun. Math. Phys.} {\bf 206} (1999) 603,
  \href{http://xxx.lanl.gov/abs/hep-th/9806099}{{\tt hep-th/9806099}};
B.~Morariu and B.~Zumino, ``Super Yang-Mills on the noncommutative torus,''
  \href{http://xxx.lanl.gov/abs/hep-th/9807198}{{\tt hep-th/9807198}};
C.~Hofman, E.~Verlinde, and G.~Zwart, ``U-duality invariance of the
  four-dimensional Born-Infeld theory,'' {\em JHEP} {\bf 10} (1998) 020,
  \href{http://xxx.lanl.gov/abs/hep-th/9808128}{{\tt hep-th/9808128}};
C.~Hofman and E.~Verlinde, ``U-duality of Born-Infeld on the noncommutative
  two-torus,'' {\em JHEP} {\bf 12} (1998) 010,
  \href{http://xxx.lanl.gov/abs/hep-th/9810116}{{\tt hep-th/9810116}};
C.~Hofman and E.~Verlinde, ``Gauge bundles and Born-Infeld on the
  noncommutative torus,'' {\em Nucl. Phys.} {\bf B547} (1999) 157,
  \href{http://xxx.lanl.gov/abs/hep-th/9810219}{{\tt hep-th/9810219}}.

\bibitem{morita}
A.~Konechny and A.~Schwarz, ``BPS states on noncommutative tori and duality,''
  {\em Nucl. Phys.} {\bf B550} (1999) 561,
  \href{http://xxx.lanl.gov/abs/hep-th/9811159}{{\tt hep-th/9811159}};
A.~Konechny and A.~Schwarz, ``Supersymmetry algebra and BPS states of super
  Yang-Mills theories on noncommutative tori,'' {\em Phys. Lett.} {\bf B453}
  (1999) 23, \href{http://xxx.lanl.gov/abs/hep-th/9901077}{{\tt
  hep-th/9901077}};
A.~Konechny and A.~Schwarz, ``1/4-BPS states on noncommutative tori,'' {\em
  JHEP} {\bf 09} (1999) 030, \href{http://xxx.lanl.gov/abs/hep-th/9907008}{{\tt
  hep-th/9907008}};
B.~Pioline and A.~Schwarz,
``Morita equivalence and T-duality (or B versus Theta),''
JHEP {\bf 9908}, 021 (1999)
[hep-th/9908019].

\bibitem{Hashimoto:1999yj}
A.~Hashimoto and N.~Itzhaki,
``On the hierarchy between non-commutative and ordinary supersymmetric  Yang-Mills,''
JHEP {\bf 9912}, 007 (1999)
[hep-th/9911057].

\bibitem{lattice}
S.~Elitzur, R.~B.~Pearson and J.~Shigemitsu,
``The Phase Structure Of Discrete Abelian Spin And Gauge Systems,''
Phys.\ Rev.\  {\bf D19}, 3698 (1979);
D.~Horn, M.~Weinstein and S.~Yankielowicz,
``Hamiltonian Approach To $\Zint_N$ Lattice Gauge Theories,''
Phys.\ Rev.\  {\bf D19}, 3715 (1979);
A.~Ukawa, P.~Windey and A.~H.~Guth,
``Dual Variables For Lattice Gauge Theories And The Phase Structure Of Z(N) Systems,''
Phys.\ Rev.\  {\bf D21}, 1013 (1980).



\bibitem{Cardy}
J.~L.~Cardy and E.~Rabinovici,
``Phase Structure Of $\Zint_p$ Models In The Presence Of A Theta Parameter,''
Nucl.\ Phys.\  {\bf B205}, 1 (1982).
J.~L.~Cardy,
``Duality And The Theta Parameter In Abelian Lattice Models,''
Nucl.\ Phys.\  {\bf B205}, 17 (1982).

\bibitem{6d}
E.~Witten,
``New `gauge' theories in six dimensions,''
JHEP {\bf 9801}, 001 (1998)
[hep-th/9710065];
B.~Kol,
``On 6d `gauge' theories with irrational theta angle,''
JHEP {\bf 9911}, 017 (1999)
[hep-th/9711017];
M.~Alishahiha and Y.~Oz,
``Supergravity and 'new' six-dimensional gauge theories,''
hep-th/0008172.

\bibitem{ncos}
R.~Gopakumar, S.~Minwalla, N.~Seiberg and A.~Strominger,
``OM theory in diverse dimensions,''
JHEP {\bf 0008}, 008 (2000)
[hep-th/0006062];
I.~R.~Klebanov and J.~Maldacena,
``1+1 dimensional NCOS and its U(N) gauge theory dual,''
hep-th/0006085.

\bibitem{Landi:1999ey}
G.~Landi, F.~Lizzi and R.~J.~Szabo,
``From large N matrices to the noncommutative torus,''
hep-th/9912130.


\bibitem{Witten:1979ey}
E.~Witten,
``Dyons Of Charge $e \theta / 2 \pi$,''
Phys.\ Lett.\  {\bf B86}, 283 (1979).



\bibitem{tHooft:1981ht}
G.~'t Hooft,
``Topology Of The Gauge Condition And New Confinement Phases In Nonabelian Gauge Theories,''
Nucl.\ Phys.\  {\bf B190}, 455 (1981).

\bibitem{cassels}
J. W. S. Cassels, 
``An introduction to Diophantine approximation'',
Haffner Pub. Co., NY (1957, reprinted 1972);
S. Lang,
``An introduction to Diophantine approximations'',
Addison-Wesley Pub. Co. (1966).

\bibitem{Hashimoto:1999ut}
A.~Hashimoto and N.~Itzhaki,
``Non-commutative Yang-Mills and the AdS/CFT correspondence,''
Phys.\ Lett.\  {\bf B465}, 142 (1999)
[hep-th/9907166].

\bibitem{Maldacena:1999mh}
J.~M.~Maldacena and J.~G.~Russo,
``Large N limit of non-commutative gauge theories,''
JHEP {\bf 9909}, 025 (1999)
[hep-th/9908134].

\bibitem{Barbon}
J.L.F. Barb\'on, private communication.



\bibitem{Giveon:1994fu}
A.~Giveon, M.~Porrati and E.~Rabinovici,
``Target space duality in string theory,''
Phys.\ Rept.\  {\bf 244}, 77 (1994)
[hep-th/9401139].

\bibitem{GL}
R.~Gregory and R.~Laflamme,
``Black strings and p-branes are unstable,''
Phys.\ Rev.\ Lett.\  {\bf 70}, 2837 (1993)
[hep-th/9301052];
R.~Gregory and R.~Laflamme,
``The Instability of charged black strings and p-branes,''
Nucl.\ Phys.\  {\bf B428}, 399 (1994)
[hep-th/9404071].

\bibitem{Kogan}
J.~L.~Barbon, I.~I.~Kogan and E.~Rabinovici,
``On stringy thresholds in SYM/AdS thermodynamics,''
Nucl.\ Phys.\  {\bf B544}, 104 (1999)
[hep-th/9809033];
S.~A.~Abel, J.~L.~Barbon, I.~I.~Kogan and E.~Rabinovici,
``String thermodynamics in D-brane backgrounds,''
JHEP {\bf 9904}, 015 (1999)
[hep-th/9902058];
S.~A.~Abel, J.~L.~Barbon, I.~I.~Kogan and E.~Rabinovici,
``Some thermodynamical aspects of string theory,''
hep-th/9911004;
E.~Martinec and V.~Sahakian,
Phys.\ Rev.\  {\bf D59}, 124005 (1999)
[hep-th/9810224].

\bibitem{Mandelbrot}
B. Mandelbrot, {\it 
Multifractals and $1/F$ Noise : Wild Self-Affinity in Physics},
selected works (1963-1976),
Springer Verlag, 1999.

\end{thebibliography}
\end{document}